\documentclass[12pt,tightenlines,eqsecnum,floats,aps,amsmath,amssymb,nofootinbib,prd,showpacs]{revtex4}

\usepackage{setspace}
\usepackage{amsmath,amssymb,amsfonts,amsthm}
\usepackage{graphicx}
\usepackage{enumerate} 
\usepackage{colordvi} 

\def\be{\begin{equation}}
\def\ee{\end{equation}}
\def\ba{\begin{eqnarray}}
\def\ea{\end{eqnarray}}

\def\H{{\cal H}}

\def\Hk{\H_{\rm kin}}
\def\Hp{\H_{\rm phy}}

\def\a{\alpha}
\def\b{\beta}

\def\lp{{\ell}_{\rm Pl}}

\def\comp{\mathbb{C}}

\def\b{$\bullet\,\, $}

\def\a{\mathfrak{a}}

\def\WDW{\rm WDW\,\,}
\def\MWDW{\rm MWDW\,\,}

\usepackage{enumerate}

\usepackage{colordvi}
\newcounter{mnotecount}[section]

\newcommand{\comment}[1]{}

\begin{document}

\preprint{\vbox{\baselineskip=12pt \rightline{IGPG-07/01-02}
\rightline{NSF-KITP-07-05 }}}

\title{Loop Quantum Gravity:\\ Four Recent Advances and a Dozen
Frequently Asked Questions}

\author{Abhay Ashtekar${}^{1,2}$}
\email{ashtekar@gravity.psu.edu} \affiliation{${}^{1}$Institute for
Gravitational Physics and Geometry, Physics Department, Penn
State, University Park, PA 16802, U.S.A.}

\begin{abstract}

As per organizers' request, my talk at the 11th Marcel Grossmann
Conference consisted of two parts. In the first, I illustrated
recent advances in loop quantum gravity through examples. In the
second, I presented an overall assessment of the status of the
program by addressing some frequently asked questions. This
account is addressed primarily to researchers outside the loop
quantum gravity community.

\end{abstract}


\maketitle

\section{Examples of Recent Advances}
\label{s1}

In this section I will attempt to provide a flavor of the progress
that is being made on various fronts of Loop Quantum Gravity
(LQG).  I have chosen examples from four areas:\\ i) Mathematical
foundation of the theory \cite{lost};\, ii) Planck scale physics
near the big bang singularity in the FRW cosmologies \cite{aps};\,
iii) Effective matter theories obtained by integrating out
gravitational degrees of freedom \cite{fl};\, and iv) Recovery of
the graviton propagator starting from a non-perturbative,
background independent theory \cite{propagator}. The first two of
these developments arose in the canonical formulation of the
theory while the last two refer to spin foams ---the path integral
framework.

I want to emphasize that these are only illustrations. Because of
time limitation, I could not include other recent advances, in
particular the mathematically interesting extensions of gauge
theories using quantum groups \cite{ao};\, the master constraint
program \cite{master} and algebraic quantum gravity \cite{aqg};\,
explorations of the Planck scale geometry in symmetry reduced
midi-superspaces \cite{madrid};\, ramifications of quantum
geometry for dynamics of matter \cite{matter};\, ideas on black
hole evaporation and information loss \cite{ab}\, and numerous
phenomenological developments in quantum cosmology \cite{mbrev}.
For detailed reviews, see, e.g., \cite{alrev,crbook,ttbook}.

\subsection{Power of background independence: uniqueness of LQG kinematics}
\label{s1.1}

In the Hamiltonian framework of any background independent theory
---such as general relativity--- one encounters first class
constraints which tell us that diffeomorphisms generate gauge
transformations.%
\footnote{For simplicity of presentation, in this discussion I
will restrict myself to spatially compact space-times. In the
asymptotically flat (or AdS) context, constraints generate only
those diffeomorphisms which are asymptotically identity; only
these are treated as gauge. Although for definiteness I have
discussed the Dirac quantization program below, considerations
remain unaltered in the BRST scheme.}
In the fifties and sixties Bergmann and Dirac analyzed such
classical systems and Dirac introduced a systematic quantization
program. Here, one first ignores constraints and introduces a
kinematic framework consisting of an algebra $\a$ of quantum
operators and a representation thereof on a Hilbert space $\Hk$.
This provides the arena for defining and solving the quantum
constraints. When equipped with a suitable inner product, the space
of solutions defines the physical Hilbert space $\Hp$.

In particle mechanics one often begins with the Heisenberg-algebra
generated by operators ${q}$ and ${p}$, satisfying the canonical
commutation relations, or the Weyl algebra generated by
${U}(\lambda) := \exp i\lambda q$ and $V(\mu):= \exp i\mu p$. The
von Neumann theorem ensures us that, under suitable physically
motivated assumptions, the algebra admits a unique irreducible
representation, namely the standard Schr\"odinger one. Therefore for
constrained mechanical systems one generally uses this
representation for quantum kinematics. However for field theories
---i.e. for systems with an infinite number of freedom--- the
uniqueness theorem fails and there is an infinite number of
inequivalent representations. To select a preferred one, additional
physical inputs are necessary.

A systematic approach to finding these representations is provided
by the celebrated Gel'fand-Naimark-Segal (GNS) construction which
may be summarized as follows. A \emph{state} on a $\star$-algebra
$\a$ is a positive linear functional, i.e., a linear map $F$ from
$\a$ to $\comp$ such that $F(a^\star a) \ge 0$ for all $a$ in $\a$
and $F(I)$=1 (where $I$ is the identity element of $\a$). Given a
state $F$, GNS provided a constructive procedure to obtain a
$\star$-representation of $\a$ by operators on a Hilbert space
$\H$ such that: i) there is a normalized cyclic vector $\Psi_F$ in
$\H$ (i.e., the action $\a$ on $\Psi_F$ yields a dense sub-space
of $\H$); and, ii) $F(a) = \langle \Psi_F|a|\Psi_F\rangle$. Thus,
in this GNS representation the original positive linear functional
yields just expectation values of operators in the cyclic state.%
\footnote{In the mathematical literature, $\Psi_F$ is often
referred to as the `vacuum' although it may not be directly
related to the Hamiltonian of the theory.}
To summarize, the task of finding a representation is \emph{neatly
reduced to that of finding a state $F$ on the algebra.} If the
state $F$ is invariant under an automorphism of $\a$, that
automorphism is represented by an unitary transformation on $\H$.
For free field theories in Minkowski space-time, the requirement
of Poincar\'e invariance (together with certain technical
conditions) selects a unique state $F$ on the Weyl algebra of
operators ---the corresponding $\Psi_F$ is the Fock vacuum. Since
$\Psi_F$ is Poincar\'e invariant, the Poncar\'e group is unitarily
implemented on the Fock space $\H$.

In LQG, the basic variables are a gravitational spin connection
$A_a^i$ on a 3-dimensional manifold $M$ and its conjugate momentum
$E^a_i$ which enjoys the geometric interpretation of an
orthonormal triad on $M$ (with density weight 1). The
$\star$-algebra $\a$ is generated by holonomies $h_e$ of $A_a^i$
along edges $e$ in $M$ and fluxes $E_{f,S} := \int_S E^a_i f^i
dS_a$ of triads (smeared by test fields $f^i$) across 2-surfaces
$S$ in $M$. (It is thus analogous to the algebra generated by
functions $\exp i\lambda q$ and $p$ on the phase space of a
non-relativistic particle.) However, since the system has an
infinite number of degrees of freedom, $\a$ admits infinitely many
inequivalent representations. The GNS construction provides a
convenient avenue to arrive at a preferred representation.

What condition should we use to select a state $F$ or a class of
such states? The construction of the algebra does not require a
background geometry or indeed any background field. As a
consequence, each diffeomorphisms on $M$ gives rise to an
automorphism of this algebra. To promote background independence
in the quantum theory it is natural to ask that the state be
invariant under these automorphisms. A surprisingly powerful
recent theorem is that the algebra $\a$ admits \emph{exactly one}
diffeomorphism invariant state \cite{lost}! In this precise sense,
\emph{quantum kinematics of LQG is uniquely determined by the
requirement of background independence.} Note that in contrast to
the situation with Poincar\'e invariance, there is no assumption
about dynamics here; the requirement of diffeomorphism invariance
is much more powerful than that of Poincar\'e invariance.

The GNS representation that results from this $F$ has been known
for sometime now. It was introduced ab-initio some ten years ago
\cite{aljb} and underlies the quantum geometry of LQG kinematics.
Specifically, it is in this representation that one can introduce
a spin network basis \cite{spinnet} and show that geometric
operators such as areas of surfaces and volumes of regions have
purely discrete eigenvalues \cite{alrev,crbook,ttbook}. The
uniqueness result shows that the origin of these features lies in
background independence and makes these results compelling.

\subsection{Quantum nature of the big bang in FRW cosmologies}
\label{s1.2}

In the standard cosmology based on classical general relativity,
space-time and matter are both born at the big bang and it is
meaningless to ask what was there before. However, one expects
quantum effects to dominate when the curvature enters the Planck
regime. Thus, big-bang is a prediction of general relativity in a
domain where it is inapplicable! Key questions to any quantum
gravity theory are then: What is the quantum nature of the big
bang? Is the classical singularity resolved by quantum effects? Is
there a quantum extension of the classical space-time?

These questions were raised already in the late sixties. To analyze
them, DeWitt \cite{bd} and Misner \cite{cm} observed that in
classical general relativity one can restrict oneself just to
spatially homogeneous and isotropic situations to an excellent
degree of approximation and suggested that one begin with the same
strategy also in quantum theory. This led to quantum cosmology. Now,
imposition of these symmetries freezes all but a finite number of
degrees of freedom. Therefore, field theoretical difficulties are
by-passed and one is led to a quantum mechanical system. Recall
however that background independence of general relativity leads to
constraints. Because of strong spatial symmetries most of them are
automatically satisfied. However, one Hamiltonian constraint
survives. Thus we are led to a quantum mechanical system with a
single constraint. Since there are only a finite number of degrees
of freedom, it was customary to represent kinematical quantum states
by wave functions $\Psi(a,\phi_i)$ of the scale factor $a$ and
matter fields $\phi_i$. Imposition of the quantum constraint then
reduces to solving a second order differential equation, called the
Wheeler-DeWitt (\WDW) equation, which governs quantum dynamics.
However, because of background independence, extracting physics is
not trivial: a priori, there is no time, no obvious inner product on
solutions to this equation nor candidate Dirac observables.
Therefore, much of the early work was confined to a WKB
approximations.

However, in the case when matter fields include a massless scalar
field $\phi_o$ , it is possible to complete the
Misner-Wheeler-DeWitt (\MWDW) program \cite{aps}. The massless
scalar field serves as an intrinsic or emergent time variable both
in classical and quantum theories. The \WDW equation takes the
form
\be \label{wdw} \partial_{\phi_o}^2 \Psi(a,\phi_i) =
\underline\Theta \Psi(\phi_i,a) \ee
where $\Theta$ is a second order differential operator involving
only the scale factor and matter fields other than $\phi_o$. one
can construct the Hilbert space of physical states by a group
averaging method used in full LQG and find a convenient, complete
set of Dirac observables. One can now return to the key questions
posed at the beginning of this sub-section. Thus, one can start
with a quantum state which is peaked, in a well-defined sense, at
a chosen FRW solution at late times and evolve it backward. The
good news is that it remains peaked at the classical trajectory
showing that general relativity is recovered from this quantum
framework. The bad news is that the agreement continues all the
way into the big bang singularity. Thus, in the \MWDW theory, the
singularity is \emph{not} resolved by quantum effects.

Because there are only a finite number of degrees of freedom, it
appeared for about two decades that the big bang singularity
cannot be resolved in quantum cosmology without an external input
such as matter fields violating energy conditions, or new boundary
conditions, e.g., a la Hartle and Hawking. However, six years ago,
Bojowald \cite{mb} showed that this status-quo changes
dramatically in loop quantum cosmology. The primary reason is that
if one closely mimics full LQG in the symmetry reduced models, one
is led to a representation which is inequivalent to
Schr\"odinger's. Thus loop quantum cosmology (LQC) is
\emph{inequivalent to the \MWDW theory already at the kinematical
level.}

Last year, the program was completed  by first constructing the
physical Hilbert and Dirac observables and then analyzing quantum
dynamics in detail \cite{aps}. Now, the second order differential
operator $\underline\Theta$ of the \WDW equation (\ref{wdw}) is
replaced by a second order \emph{difference} operator $\Theta$. In
the low curvature region, $\Theta$ is well approximated by
$\underline\Theta$ but near the Planck regime there are major
differences. In particular, if we start with a state which is
semi-classical at late times and evolve backwards, it follows the
classical trajectory only until scalar curvature $R$ ---the only
independent curvature invariant for this model--- reaches a
critical value $R_{\rm crit} \approx 13\pi/{\ell_{\rm Pl}}^2$.
Then quantum geometry gives rise to an effective repulsive force
so that the wave function bounces, giving rise to a pre-big-bang
branch which again obeys the classical Einstein equations once the
curvature becomes significantly less than $R_{\rm crit}$.

This quantum evolution has been analyzed in detail using numerical
methods as well as effective equation techniques for the $k=0$ and
$k=1$ FRW models with and without cosmological constant (and with
less rigor and completeness for the $k=-1$ and the Bianchi I
anisotropic models). I will conclude with a few illustrative
examples from the $k=1$ case. Here, Einstein's equations imply that
the universe expands after the big bang to a maximum radius $a_{\rm
max}$ and then recollapses. First, one can ask for the volume
$V_{\rm min}$ of the universe at the quantum bounce. This depends on
the classical trajectory about which the wave function is peaked at
a late time. If the trajectory is such that $a_{\max}$ is a
megaparsec, then $V_{\rm min} \approx  6 \times 10^{16} {\rm cm}^3$,
some $10^{115}$ times larger than the Planck volume! Furthermore
$V_{\rm min}$ scales as $a_{\rm max}^2$, whence more `macroscopic'
the universe, larger the value of $V_{\rm min}$. Quantum effects
dominate when the space-time curvature or matter density
---and not of volume--- enters the Planck regime. Second, as a
result of the recollapse, in the $k=1$ case the classical universe
exhibits a big bang as well as a big crunch singularity. Both are
replaced by bounces in the quantum theory. Therefore, we are led
to a cyclic scenario. Thus, quantum space-time is \emph{enormously
larger} than what classical general relativity would have us
believe. Finally, one of the outstanding questions in full LQG is
whether the theory admits a rich semi-classical sector. Even
though the LQC departures from general relativity are small when
$\rho \ll \rho_{\rm Pl}$, the expansion lasts for a very long
time. Can the quantum effects accumulate so that the universe does
not recollapse or recollapses at a value significantly different
from $a_{\rm max}$? Indeed, using an early form of the LQC
Hamiltonian constraint, Green and Unruh \cite{gu} argued that
there would be no recollapse. However the recent, much more
complete and detailed analysis \cite{aps} has shown that the
universe does recollapse in LQC and agreement with classical
general relativity on $a_{\rm max}$ is excellent. Even for
universes which are so small that $a_{\max} \approx 30 \lp$, the
classical Friedmann formula $\rho_{\min} = 3/(8\pi\, G
a_{\max}^2)$ holds to one part in $10^{-5}$ and the agreement
improves greatly for `macroscopic' universes, i.e., ones with
macroscopic values of $a_{\rm max}$. This detailed agreement
suggests that LQG has the potential for curing the ultraviolet
divergences of general relativity while admitting a rich
semi-classical sector.

These attractive results have been obtained for the simplest
models ---FRW space-times with a massless scalar field--- by
making a symmetry reduction at the classical level and then going
to the quantum theory. This LQC is yet to be derived from full
LQG. Within this key limitation, it is nonetheless interesting
that one can probe Planck scale physics near the most interesting
curvature singularities, the results are strikingly different from
the \MWDW theory, and the difference originates in the quantum
nature of geometry underlying LQG.

\subsection{Emergence of non-commutative quantum field theory}
\label{s1.3}

An outstanding question facing any quantum gravity theory is: What
are the ramifications of the quantum nature of gravity on dynamics
of ordinary matter fields? In the canonical framework, analysis
involves two steps: i) construction of quantum field theories on
quantum geometries; and ii) dynamical selection of suitable
quantum geometries appropriate for low energy physics. There has
been considerable work on the first part which has clarified in
particular how the fundamental discreteness inherent to quantum
geometry could tame the ultra-violet divergences that plague
quantum field theories in the continuum \cite{matter,ttbook,st}.
However, investigation of the second step is still at a
preliminary stage. Consequently, reliable predictions on quantum
gravity ramifications for low energy physics are yet to emerge
from the canonical theory.%
\footnote{See, however, the work of the Madrid group \cite{madrid}
based on Hamiltonian methods. Although it does not lie in the
strict confines of canonical LQG, it is based on quantum geometry
considerations.}

Path integral methods are better suited to address this issue. For,
the program can be stated rather simply: Given a quantum theory with
matter, what is the effective theory that results after
gravitational degrees of freedom have been integrated out? The
technical execution of the program is however not as simple because
of the task of integrating out the gravitational degrees of freedom.
What measure should one use? Since Einstein gravity is not
perturbatively renormalizable, one cannot use methods that are
readily available in the more familiar Minkowskian field theories.

Spin foams provide a non-perturbative and rigorous framework to
evaluate path integrals for quantum gravity \cite{spinfoams}.
Their strength lies in the fact that they supply a definition of a
quantum space-time in algebraic and combinatorial terms. In
essence, one can think of the quantum space-time as a generalized,
2-dimensional Feynman diagram with degrees of freedom propagating
along surfaces. These considerations can be made explicit in 2+1
dimensions. Through a series of steps, it has recently become
possible to go further and integrate out the gravitational degrees
of freedom to obtain effective theories for matter fields
\cite{fl}. Furthermore, the process seems to naturally lead to
non-commutative field theories where particles can be envisaged as
living in a \emph{non-commutative space-time.}

Let us start with the partition function $Z = \int\, Dg\,\int\,
D\phi\,\, \exp i(S[\phi, g] + S[g])$ where $S[\phi,g]$ is the
action for a scalar field $\phi$ (with non-derivative
interactions) on a 3-dimensional space-time with metric $g_{ab}$
and $S[g]$ is the general relativity action. The goal is to
integrate out the gravitational degrees of freedom and obtain an
effective action $S_{\rm eff}[\phi]$ for the matter field $\phi$.
The idea is to achieve this by first expanding out the
$\phi$-integral in terms of Feynman diagrams which depend on a
background metric $g_{ab}$:\, $Z = \sum_{\Gamma}\, C_\Gamma \,
\left(\int Dg\, I_\Gamma [g]\, \exp iS[g]\right) \equiv
\sum_\Gamma C_\Gamma\, \tilde{I}_\Gamma$. The next and even more
non-trivial step is to re-sum the Feynman diagrams and obtain an
effective action: $Z = \int D\phi\, \exp iS_{\rm eff}[\phi, g_o]$,
where $g_o$ is the flat metric.

This procedure has been carried out by making an ansatz for
coupling gravity to the Feynman diagrams of particles \cite{fl}. A
rigorous derivation of the ansatz starting from the action
$S[\phi, g]$ is still lacking. But the final result is rather
simple and therefore attractive: The standard Abelian products of
fields in the original action $S[\phi,\, g]$ are just replaced by
a non-Abelian $\star$-product in $S_{\rm eff}[\phi,\, g_o]$. If
metrics $g_{ab}$ have signature +,+,+, the resulting theory is
called Riemannian while if the signature is +,-,-, (or -,+,+) it
is the Lorentzian theory of direct physical interest. Detailed
analysis has been carried out in both cases. (Note that the
partition function involves $\exp iS$ in both cases; thus the
traditional Euclidean quantum theory is not considered.)

While the action $S[\phi,g_o]$ is invariant under the six
dimensional Poincar\'e group, $S_{\rm eff}[\phi, g_o]$ is
invariant under the so-called $\kappa$ deformation of the
Poincar\'e group ---which is again six dimensional--- where
$\kappa$ now has the value $4\pi G$. Ordinarily, deformed
Poincar\'e theories suffer from huge ambiguities because we do not
a priori know which elements of the Hopf algebra are to be
identified with physical energy and momentum. However, in the
present case these are resolved because the deformation is tied
directly to quantum gravity effects stemming from the
Ponzano-Regge model. The effect of quantum gravity on matter is
two folds. First, the mass gets renormalized via $m\rightarrow
\sin \kappa m/\kappa$. Second, the momentum space is no longer a
flat space but the homogeneous space SO(3) in the Riemannian case
and SO(2,1) in the Lorentzian. The underlying group structure
enables one to `add' momenta of particles but this operation is
non-Abelian. These particles can be regarded as moving in the
`dual', non-commutative position space co-ordinatized by $X_i$
satisfying the relations:
\be [X_i, X_j] = i\kappa \hbar \epsilon_{ijk}X_k\, . \ee
The $\star$-algebra generated by the three operators $X_i$ subject
to these commutation relations is referred to as the
$\kappa$-Minkowski space. Detailed expressions show that energy of
matter particles is bounded above by $1/\kappa$ in any rest frame
(because $p_0^2 -p_1^2-p_2^2< 1/\kappa^2$). Similarly, there is a
minimal length scale $\sim \kappa\hbar$ accessible to the theory.

Thus, thanks to the spin foam technology that has been
systematically developed over the last decade, a new approach to
analyze the quantum gravity effects on matter fields has now
emerged. Although some issues still remain, it is encouraging that
the analysis provides a concrete realization of several features
that have been heuristically expected in the literature on
physical grounds: a quantum gravity induced cut-off,
non-commutative fields, effective non-commutative space-times, and
the idea proposed by Snyder \cite{snyder} (sixty years ago!) that
a curved momentum space could regularize field theories. However,
the detailed work to date \emph{is} tied to features of 3d
gravity. Although extensions to 4 dimensions have already been
suggested, it is not yet clear which of these features will filter
down in 4-dimensional effective theories.

\subsection{The graviton propagator}
\label{s1.4}

Background dependence of LQG has powerful consequences. As we saw in
section \ref{s1.1}, it selects a unique kinematical framework and
leads one to a specific quantum theory of geometry with an in-built
discreteness \cite{lost,aljb}. This in turn has implications on the
quantum nature of singularities \cite{mb,aps} and the ultraviolet
behavior of matter fields \cite{matter,ttbook}. However, precisely
because of this emphasis on Planck scale discreteness, familiar
properties of gravity and matter that can be easily derived using
space-time continuum are now difficult to obtain. For instance in
the traditional perturbative framework, the gravitational attraction
between two point masses arises from an exchange of virtual
gravitons, described by the Feynman propagator. Loop quantum
gravity, on the other hand, has no background metric. Therefore one
cannot even begin to calculate the propagator along these
traditional lines.

The tension can be illustrated by the following simple argument.
In Minkowskian field theories, the propagator is a 2-point
function $W(x,y)$ which can be expressed as a path integral
$W(x,x') = \int D\phi\, \phi(x) \phi(x') \exp iS[\phi]$. However,
this expression cannot be taken over naively to background
independent theories. For, in such a theory the measure and the
action are diffeomorphism invariant whence we would have $W(x,x')
= W(y,y')$ for any $y, y'$ which are images of $(x,x')$ under some
diffeomorphism. A distribution $W(x,x')$ with this property would
not look anything like the familiar graviton propagator which
falls off as $|x-x'|^{-2}$. Thus, the naive calculation sketched
above is inadequate. To obtain the standard propagator, at the
very least we must inject into the calculation enough structure
that lets us speak of distance between $x$ and $x'$. It turns out
that an appropriate strategy is to consider a space-time region
with a boundary on which $x$ and $x'$ lie, and fix a state on the
boundary which has enough information to speak of the
bulk-distance between these points. To obtain the propagator, one
can then integrate over fields in the bulk (i.e. interior) which
are compatible with
the given boundary state.%
\footnote{For details on why this is a natural generalization of the
standard procedure in Minkowski space-time, see \cite{propagator}
and references therein.}
First steps in the implementation of this idea have been carried out
over the last year. So far, calculations have been done only in the
Riemannian (rather than Lorentzian) framework \cite{propagator}.

Consider then $\mathbb{R}^4$ and introduce in it a topological
3-sphere $S$. Introduce on $S$ Cauchy data $(q_{ab}, k^{ab})$
induced by a flat Euclidean 4-metric. By `evolving' this data via
Einstein's equation one would recover an Euclidean 4-metric
$g_{ab}$ in the interior of $S$. However, this 4-metric will be
unique only up to diffeomorphisms (which are identity on the
boundary $S$). But the geodesic distance between $x$ and $x'$ is
invariant with respect to this diffeomorphism freedom. The idea is
that the desired propagator should be obtained by fixing points
$x,x'$ on the boundary $S$ and summing over all configurations in
the interior bulk which agree with the boundary data $(q_{ab},
k^{ab})$. In quantum theory then one fixes a LQG boundary state
$\Psi_{q,k}(s)$ peaked at some flat space initial data $(q_{ab},
k^{ab})$, where $s$ is a (diffeomorphism equivalence class of)
spin network(s) on the 3-sphere $S$. Formally the gravitational
propagator is given by
\be W^{aba'b'}(\{S,q,k;\, x,x'\})\, = \, N \sum_{s,s'}\, W[s']
\langle s'|h^{ab}(x) h^{a'b'}(x')|s\rangle \, \Psi_{q,k}[s] \ee
where $\{S,q,k; \, x,x'\}$ stands for a diffeomorphism class of
$(S, q_{ab}, k^{ab},x,x')$ and $W(s')$ the amplitude obtained by
summing over all spin-foams in the interior which are compatible
with the
spin network $s$ on the boundary.%
\footnote{ $W(s')$ is the spin-foam analog of the sum over
geometries $W({}^3g) := \int D[g]\, \exp iS_{\rm EH}[g]$ where the
integral is performed over all 4-geometries $g$ which agree with
the pre-specified 3-geometry ${}^3g$ on the boundary.}
Note that while the construction does require a state peaked at some
initial data for Euclidean space, the answer depends only on the
diffeomorphism equivalence class $\{S,q,k; \, x,x'\}$ of the quintet
inside the curly brackets.

Technically, the non-trivial parts of the calculation are of
course the evaluation of the transition amplitude $W(s)$ and the
computation of the sum. The well-developed spin-foam technology
enables one to perform the first task rigorously. The result is
manifestly finite. Computation of the sum turns out to be more
delicate. Here, there is a conceptually important but
mathematically delicate cancellation between phase factors that
arise from $W(s)$ and the semi-classical state $\Psi_{q,k}(s)$
which makes the sum convergent. This subtle interplay provides
considerable support for the way the problem has been set up. The
final result can be expressed as a power series in $\lp/R$, where
$R$ is the geodesic distance between $x$ and $x'$ with respect to
the Euclidean metric. The leading term is independent of $\lp$ and
proportional to $1/R^2$ just as one would expect from perturbative
treatments. The full result includes corrections of the order
$\lp/R$ and higher, which encode the ultraviolet non-trivialities
of LQG.

I have sketched only the main idea. There are several unresolved
issues, both conceptual and technical. Conceptually, so far all
calculations are incomplete because the required boundary state is
introduced by hand. While states used \emph{are} well motivated
the choice is far from being unique and higher order corrections
depend on the choice. Ultimately this state should represent the
`Minkowskian vacuum', determined dynamically. Technically, the
most important question is whether the construction can be
extended to the physical, Lorentzian sector.  The non-trivial
achievement is that a conceptual framework has been introduced to
calculate n-point functions within a background independent
setting, thereby bridging the Planck scale quantum geometry to the
familiar continuum physics.\\

This concludes the discussion of recent advances. The kinematic
results of section \ref{s1.1} are rigorous and apply to full LQG.
The subsequent three examples refer to dynamics: The LQC analysis
summarized in section \ref{s1.2} resolves singularities of direct
physical interest; the effective non-commutative field theory of
section \ref{s1.3} provides new a approach to quantum gravity
phenomenology; and results of section \ref{s1.4} on the propagator
open up the possibility of relating LQG to the more familiar
Minkowskian perturbation theory and pin-pointing the limitations
of the latter. However, since these three examples refer to
non-perturbative \emph{dynamics}, important issues remain:
relation of LQC with LQG in section \ref{s1.2}; a systematic
derivation of the ansatz used in section \ref{s1.3}; and
determination of the boundary state in the Lorentzian sector in
\ref{s1.4}. Together, these examples illustrate that progress is
being made on several different fronts.

\section{Frequently asked questions}
 \label{s2}

The organizers of the conference noted that there has been growing
interest in loop quantum gravity, illustrated, e.g., by some
semi-popular articles and reviews written by authors who do not work
in this field, and requested that I clarify issues that puzzle
`outsiders' and respond to questions string theorists often raise.
These outside perspectives are extremely helpful because they
can bring freshness.%
\footnote{ For example, in his recent book `Road to Reality' Roger
Penrose has given such fresh perspectives on string theory.}
The LQG community very much appreciates these efforts. However, such
reviews can also put the program in a pre-conceived, conceptual
straightjacket inherited from author's own expertise, thereby
missing the spirit of the endeavor. This is not surprising because
we all know the difference between working in a field and learning
about it by reading; reading alone cannot give a deep and
encompassing perspective that arises after years of thinking about
problems and working out technical details. My assigned task was
then to try to create a bridge from LQG to the broader quantum
gravity community, particularly string theorists.

This is not straightforward particularly because, as in any healthy,
developing field, the LQG community has diverse viewpoints on some
of the key open issues. Therefore, to prepare my talk I consulted a
number of leading researchers. But my answers typically represent
only a broad consensus rather than a sharp, unanimous view. Also, at
some points I have taken the liberty to express my personal take on
the issue.

\subsection{Structure of LQG}

\b \emph{1. Uniqueness of the LQG kinematic framework rests on the
result \cite{lost} that there is a unique diffeomorphism invariant
state. But how can this be? Shouldn't there be an infinite number of
diffeomorphism invariant states in quantum gravity?}\\

As explained in section \ref{s1.1}, a state is a positive linear
functional on a $\star$-algebra. The uniqueness result refers to
the basic holonomy-flux algebra $\a$ of LQG. The
\emph{kinematical} Hilbert space $\Hk$ of LQG that results from
the GNS construction therefore admits a unique diffeomorphism
invariant ray. All vectors in the infinite dimensional
\emph{physical} Hilbert space of LQG are indeed diffeomorphism
invariant, but they are positive linear functionals on a
\emph{different} algebra, consisting of diffeomorphism invariant
operators.

This point may seem confusing. Let me therefore consider the more
familiar example of quantum geometrodynamics (QGD). Now the $\star$
algebra $\a$ is usually taken to be generated by the metric $q_{ab}$
and its conjugate momentum $p_{ab}$ (both smeared with test fields),
subject to the standard canonical commutation relations
$[q_{ab}(x),\,\,p^{cd}(y)] = -i\hbar\, \delta_{(a}^c\,\delta_{b)}^d
\delta(x,y)\, I$. One's first inclination is to expect that $\a$
admits infinitely many diffeomorphism invariant states (e.g.
$\Psi(q) = \int_M R_{ab}R^{ab} dV_q$, where $R_{ab}$ is the Ricci
tensor of $q_{ab}$). If so, the uniqueness results in LQG would seem
strange, perhaps an artifact of some hidden assumption. Let us
therefore explore this issue. Suppose we find a diffeomorphism
invariant state $F$ on this $\a$. As in section \ref{s1.1}, let us
denote by $\Psi_F$ the cyclic vector in the Hilbert space $\H$ that
results from the GNS construction. Then it follows that the
expectation values of $q_{ab}(x)$,\,\, $q_{ab}(x) q_{a^\prime
b^\prime}(x^\prime)\,\,$ etc. in the state $\Psi_F$ must be
diffeomorphism invariant distributions on $M$. But the only such
distribution is the zero distribution. It follows immediately that
$q_{ab}(x)\Psi_f =0$. An identical argument tells us that
$p^{cd}(y)\Psi_f =0$. But this is impossible as it would contradict
the canonical commutation relations. Thus, contrary to one's initial
expectation, the canonical algebra of quantum geometrodynamics does
not admit even a single diffeomorphism invariant
state.%
\footnote{The detailed version of this argument is rigorous. There
are no subtleties concerning domains of operators because the GNS
construction guarantees that the products $q(f)p(g)$ and $p(g)q(f)$
of smeared operators have a well-defined action on $\Psi_F$. The
main result holds also for the affine algebra of Klauder's
\cite{jk}.}

Following the footsteps of LQG, let us modify the kinematical
algebra appropriately. Let $\a$ be the algebra generated by
smeared operators $q(f) := \int_M q_{ab} f^{ab}(x) d^3x$ and
exponentiated operators $\exp i \int p^{ab}(x) g_{ab}(x) d^3x$
(where $f^{ab}$ is a test tensor density of weight 1 and $g_{ab}$
a test tensor field with density weight zero). Then one can indeed
find a diffeomorphism invariant state $F$ which is completely
analogous to that of LQG. In the resulting GNS representation the
operators $q(f)$ are well defined but there is no operator
corresponding to $p(g)$ because the operators $\exp i \int
p^{ab}(x) g_{ab}(x) d^3x$ fail to be weakly continuous in $g$.
This is completely analogous to the fact that in LQG operators
$h_e$ and $E_{S,f}$ are well-defined but there is no operator
corresponding to the connection because holonomy operators fail to
be continuous in the edge $e$. Next, let us consider the cyclic
state $\Psi_F$. Reasoning of the last paragraph implies that the
state is sharply peaked at the zero metric! (The same is true in
LQG). Thus $\Psi_F$ is the analog of Witten's `non-perturbative,
diffeomorphism invariant vacuum' in 2+1 dimensions. The problem
with the canonical commutation relations is bypassed because there
is no operator $p(g)$.

I hope this close similarity between LQG and the more familiar
quantum geometrodynamics provides some intuition for the uniqueness
result of \cite{lost}. These surprisingly strong restrictions bring
out the fact that background independence is a largely unfamiliar
territory that can hold many surprises. The situation in quantum
geometrodynamics may also help in making the nature of the LQG
cyclic state and the non-existence of a connection operator in the
theory less mysterious. \\ \medskip

\b \emph{2. LQG seems to start with general relativity or
supergravity. Shouldn't Einstein's equations get quantum
corrections?}\\

Yes. And they \emph{do} receive quantum corrections in LQG.

Let us first consider QED. There one begins with the classical
Maxwell-Dirac action and then proceeds with quantization. One does
not argue that the classical action must be modified because of
quantum corrections. Yet, the effective action  ---which is meant to
incorporate all quantum effects--- has all sorts of additional terms
representing quantum corrections.

The viewpoint is similar with Einstein-matter actions in LQG. The
classical theory one `quantizes' is general relativity (or
supergravity). But Einstein equations do receive quantum
corrections. In fact, one would expect the `effective action' to
exhibit \emph{non-localities} at the Planck scale (in addition to
the common non-local terms one encounters in Minkowskian quantum
field theories).

We already know the leading corrections in quantum cosmology. The
k=0 Friedmann equation $(\dot{a}/a)^2 = 8\pi G/3$ is replaced by
$(\dot{a}/a)^2 = (8\pi G/3)[1 -\rho_{\rm matter}/\rho_{\rm crit}]$
where $\rho_{\rm crit} \approx 0.8 \rho_{\rm Pl}$. The precise
form of this quantum correction leads to profound departures from
general relativity in the Planck regime.\\
\medskip

\subsection{Quantum dynamics}
\label{s2.2}

\b \emph{3. Isn't there a large number of ambiguities in the
dynamics of LQG?}\\

Yes. Indeed, these have been pointed out in many reviews (see, e.g.
\cite{alrev}) and have constituted a focal point of concern in the
LQG community. This is precisely the current incompleteness of the
mathematical framework of LQG. Ambiguities can be broadly divided
into three classes of increasing importance.

The first type corresponds to factor ordering choices which are
present also in ordinary quantum mechanics. They are partially
reduced by the requirement that the constraint should be
self-adjoint so one can pass to the physical Hilbert space by
group averaging. These ambiguities persist in symmetry reduced
models including quantum cosmology. In simple models, explicit
calculations have shown that this freedom does not change the
qualitative predictions of the theory. In particular in the FRW
models with scalar fields, the resolution of the big bang
singularity and the main features of quantum physics near the big
bang are robust \cite{aps,mb}. Therefore there is reason to hope
that the situation would be similar in the full theory.

The second and more significant source of ambiguities is
associated with the choice of representation $j$ associated with
the new edge added by the action of the Hamiltonian constraint on
$\Hk$. This does change the constraint qualitatively. In quantum
cosmology the choice $j=1/2$ leads to a second order difference
equation which reduces to the Wheeler DeWitt equation when the
curvature is low compared to the Planck scale. Vandersloot has
shown that the use of $j>1/2$ leads to a higher order difference
equation which has many more solutions \cite{higherj}. One can
argue that the extra solutions are spurious. Starting with LQG in
2+1 dimensions, Perez has given arguments to the effect that we
should only use the fundamental representation $j=1/2$ also in 3+1
dimensions \cite{higherj}. Thus, while the issue is not
definitively settled, there are several pointers indicating that
this ambiguity could be naturally resolved.

Since there is a certain similarity with lattice gauge theories,
it is instructive to compare the two situations. In lattice gauge
theories ambiguities arise in the intermediate stage (because of
irrelevant operators) but disappear in the continuum limit. In LQG
we are already in the continuum but the underlying diffeomorphism
invariance enables one to get rid of a host of ambiguities.
However, a large number of them still persist, in particular,
those associated with the choice of an initial triangulation of
the 3-manifold $M$ and treatment of the so called `non-Euclidean
part of the constraint'. This third set of ambiguities is much
more severe. In my view, the most unsatisfactory aspect of the
current status of the program is that the physical meaning and
ramifications of these ambiguities are still poorly understood.
One can invoke `naturality' and `simplicity' criterion to remove
them but without a deeper understanding of their physical meaning,
these criteria can be subjective and therefore not compelling.\\
\medskip

\b \emph{4. In the current approach, the constraint algebra can
not be verified in quantum theory. Is this not a fatal drawback?}\\

No. \emph{Logically} there is nothing wrong in: i) first imposing
the Gauss constraint; ii) defining the Diff constraint only on the
Hilbert space of gauge invariant states and solving it; and iii)
then defining the Hamiltonian constraint on the Hilbert space of
gauge and Diff invariant states and then solving it. If this
procedure leads to a theory with a rich semi-classical sector, it
would be a viable physical theory.

However, as I just discussed, there is still a large number of
poorly controlled ambiguities in the definition of the Hamiltonian
constraint. Requiring the satisfaction of the constraint algebra
could be very helpful in reducing them and to have a better chance
at arriving at a viable theory. Progress along these lines has
been made recently by Giesel and Thiemann  in that they are able
to show that for a certain class of constructions of the quantum
constraint, the expected algebra is indeed realized on
semi-classical states \cite{aqg}.\\ \medskip

\b \emph{5. Aren't these ambiguities just a reflection of the
ambiguities associated with perturbative non-renormalizability of
Einstein gravity?}\\

\emph{No!} The two have entirely different origins: The ambiguities
in LQG arise from an incompleteness of our current understanding
while the infinite ambiguities of perturbative quantum general
relativity are an expression of the inadequacy of the Gaussian fixed
point.

2+1-dimensional general relativity is also power counting
non-renormalizable but exactly soluble in LQG and spin foam
calculations have also established this elegantly in the path
integral framework \cite{2+1}. In another direction, the Madrid
group has compared and contrasted the perturbative and
non-perturbative treatments for 2+1 gravity coupled to a scalar
field \cite{madrid}. The precise and important differences in
geometry, causal structure and correlation functions that emerge
at the Planck scale appear to bear out the qualitative
expectations and hopes researchers have expressed for some time.
Thus, a priori there is no relation between perturbative and
non-perturbative ambiguities. Finally, this view is re-enforced
also by results on `causal dynamical triangulations' \cite{cdt}.

Indeed, there exist perturbatively non-renormalizable but exactly
soluble models, e.g., the Gross-Neveau model in three dimensions.
It admits a non-Gaussian fixed point (NGFP). Initially it was
thought that the correlation functions will not be tempered
distributions, i.e., will be worse behaved than those in
renormalizable field theories, reflecting the perturbative
non-renormalizability. This turned out not to be the case! By now
there is significant evidence that Einstein gravity also admits a
NGFP \cite{reuter}. Furthermore there are some qualitative
similarities on the nature of the Planck scale geometry in these
`asymptotically safe' scenarios and in spin-foams/LQG. In
particular, the effective space-time dimension
at the Planck scale turns out to be two in both theories.\\
\medskip

\b \emph{6. Will Lorentz invariance be violated in the low energy
limit of LQG dynamics?}

Let me answer in steps in part because the question has several
connotations and inequivalent precise formulations.

LQG is based on a Hamiltonian theory and sometimes field theorists
implicitly assume that this feature would automatically lead to a
Lorentz violation. This is not the case. For example, in the
canonical description of Minkowskian field theories the full
Poincar\'e group acts unitarily. Similarly, in the asymptotically
flat context the canonical phase space of general relativity does
carry a symplectic representation of the asymptotic Poincar\'e
group and the Hamiltonian generating these transformations are the
total energy-momentum and angular-momentum. Thus, by itself the
3+1 split is not an obstruction to Lorentz invariance. Finally,
sometimes quantization of area and volume in quantum kinematics is
taken to indicate Lorentz violations. This is not the case: recall
that quantization of eigenvalues of angular
momentum operators $J_i$ does not break spherical symmetry.%
\footnote{For further discussion, see \cite{crss} and
\cite{crbook,alrev}. For an early discussion on why there is no
inherent conflict between Lorentz invariance and discreteness, see
\cite{snyder}.}

In full non-perturbative quantum gravity there is no background
metric whence some care is needed to speak of Lorentz invariance.
The question can only refer either to asymptotic symmetries in the
asymptotically flat context or effective low energy descriptions. I
would expect LQG will have the first type of Lorentz invariance
generated by global charges corresponding to asymptotic symmetries.
But unfortunately so far global issues related to asymptotic
flatness have received very little attention.

For effective low energy theories, the main issue is whether the
effective actions will have terms which will violate \emph{local}
Lorentz invariance. For instance the manner in which quantization
ambiguities are resolved may require a background structure and
this in turn may give rise to background fields in the effective
low energy theory, violating Lorentz invariance. Such a violation
could rule out the theory because there are very strict
experimental constraints and also theoretical arguments which say
that the Lorentz violations at very high energies would trickle
down to low energy physics because of closed loop effects in the
Feynman expansion. However, these constraints arise by assuming
that the violation is due to a background vector field (a rest
frame) or a background tensor field. While quantum geometry of LQG
probably undergoes violent fluctuations at the Planck scale, it
seems highly unlikely that their coarse graining will give rise to
such background fields. Their effect is likely to be analogous to
the 2+1 effective field theory \cite{fl} I discussed in section
\ref{s1.3}. That description is free of all the usual experimental
or theoretical difficulties \cite{lv} normally discussed under
`Lorentz violations'. Indeed, on one particle states, the action
of the Lorentz group is the standard one. However, in the
effective theory, spin statistics and addition of momenta is
non-standard. These modifications are not what is normally called
`Lorentz violations'. Experimental constraints on them appear not
to have been discussed in the literature.

However, it \emph{is} true that some choices of quantization of
operators in the Hamiltonian constraints may lead to Lorentz
violations in the effective theory in the standard sense, leading to
conflicts with experiments. So, the requirement that there be no
such violations will serve as an important criterion in narrowing
down the ambiguities.%
\footnote{This strategy was explicitly realized in a recent
analysis of the CGHS model from a canonical quantization
perspective in the recent work of Tavares, Varadarajan and the
author.}
Such external, physical criteria are essential since a theory with
a large number of free parameters will not have predictive power
in practice.

\subsection{Spin foams, black hole entropy and quantum cosmology}

\b \emph{7. Much of the work in spin foams is carried out in the
Riemannian context which has no simple relation to the physical
Lorentzian sector. Why is it then interesting?}\\

Because it addresses some of the long standing problems in a
background independent fashion, with all due care of mathematical
physics. Faced simultaneously with a number of difficult technical
and conceptual issues, it is not uncommon in mathematical physics to
ignore one or two important features in order to gain insight into
the remaining aspects of the problem. An outstanding example is the
AdS/CFT correspondence in which one makes certain aspects of quantum
gravity mathematically tractable at the cost of using unphysical
boundary conditions in which not only there is a negative
cosmological constant but the extra compact dimensions have radius
of a cosmological ---rather than Planck--- size. Furthermore, unlike
the AdS/CFT correspondence, the Riemannian spin-foams can directly
suggest strategies for the physically interesting Lorentzian sector.
As we saw in section \ref{s1.3} in 3 dimensions these strategies can
be generally implemented in detail. In 4 dimensions, arguments
generally go through at a level of rigor that is often considered
acceptable in particle physics.%
\footnote{In this discussion, I restricted myself to more
conceptual issues. A number of more technical points were
addressed by Laurent Friedel in
http://www.math.columbia.edu/~woit/wordpress/?p=330\,\, January
23rd, two entries under \emph{L says} towards the end. }
\\\medskip

\b \emph{8. In the black hole entropy calculation, what is the
justification of assuming the Boltzmann statistics for punctures?}\\

This is a misconception! No such assumption was made. To compute
entropy, one just counts the number of Chern Simons states on the
horizon ---i.e., the number of states of the quantum horizon
geometry---  paying due respect to the subtleties of diffeomorphism
invariance. In this calculation, the punctures are not treated as
particles, whence the issue of their statistics does not even arise.
At the end of the calculation, one may try to reinterpret the result
by constructing a heuristic picture of the black hole horizon as a
gas of punctures. One can then conclude that one would reproduce the
result of the calculation by using Boltzmann statistics for the
hypothetical puncture particles.

Perhaps an analogy would help. From the Schwarzschild solution we
know that the radius $R_H$ of the horizon of a black hole of $M$ is
given by $R_H = 2GM/c^2$. There are also celebrated, 18th century
calculations by Mitchell and Laplace which use the formula for
escape velocity from Newtonian gravity, set the escape velocity to
the speed of light and conclude that the radius $R$ of a black hole
of mass $M$ is given by $R= 2GM/c^2$. The frequently asked question
I began with is somewhat analogous to asking to Schwarzschild:
``what is the justification of setting the escape velocity to be $c$
when the speed of light is not absolute in Newtonian gravity?''
Concepts used explicitly and implicitly in the question are irrelevant
to the systematic derivation.\\
\medskip

\b \emph{9a. A puzzle about black hole entropy is that it scales
with the area of the black hole, not the volume. While LQG
calculations directly use the physical, black hole space-time in
contrast to stringy discussions, isn't it the case that they do not
explain why entropy does not scale with the volume?}\\

The intuition about scaling with volume comes from ordinary
thermodynamical systems. But because of the singularity, the
notion of `volume of a black hole' does not have well-defined
meaning. Let us first consider a static star. Its volume at an
instant of time can be calculated by measuring the volume of the
3-surface which is orthogonal to the Killing field and whose
boundary is the surface of the star at that instant. In the case
of a static black hole, the Killing field is \emph{space-like} in
the interior of the horizon whence the 3-surface orthogonal to it
is \emph{time-like}. If one just takes a cross-section of the
horizon and asks how much volume it contains, one can get any
answer between zero and infinity, depending on the choice of the
3-surface whose boundary is the given cross-section (and the
singularity). So it is meaningless to look for an expression of
entropy that scales with volume. \\
\medskip

\b \emph{9b. If one is not considering black-hole interior, what is
then one counting?}\\

The viewpoint in LQG is that entropy is not an intrinsic attribute
of space-time but depends on its division into exterior and interior
regions. Operationally, it is tied to the class of observers who
live in the exterior region, for whom the isolated horizon is a
\emph{physical} boundary that separates the part of the space-time
they can access from the part they can not. (This point is
especially transparent for cosmological horizons which are also
encompassed by the LQG calculation and for which there is no
intrinsic analog of the `black hole region'.)

While there is an `observer dependence' in this sense, entropy
cannot refer to all the interior degrees of freedom that are
inaccessible to the observers under consideration: Since inside the
horizon one can join-on entire universes which do not communicate to
the exterior region, the number of potential interior states
compatible with the data accessible to the exterior observers is
uncontrollably large. Instead, as in the membrane paradigm, in LQG
one counts those black hole states which can interact with the
outside world, whence the entropy refers to the micro-states of the
boundary itself, i.e. of the quantum geometry of the horizon. (For
further discussion, see sections VI.C and VII of \cite{abk}.)

To summarize the goal of LQG calculations has been to answer the
following question: Given that there is an isolated horizon, what is
the entropy associated with it? Because of the conditional nature of
this question, one begins with a suitably restricted sector of
general relativity and \textit{then} carries out quantization.
However, as a result, the description is only an effective one.
Fortunately, for thermodynamic considerations involving large black
holes, effective descriptions are adequate. \\
\medskip

\b \emph{10. Isn't Loop Quantum Cosmology too restrictive because of
the huge symmetry reduction? }

Absolutely! Inclusion of inhomogeneities is crucial. This is a focal
point of significant current research.

However, it is quite possible that the qualitative results are more
robust than one has any right to expect a priori. This possibility
is suggested by two considerations. First, the
Belinskii-Lifshitz-Khalatnikov  conjecture
---which seemed too sweeping to be true when it was first proposed
some 30 years ago--- has now received considerable support in
classical general relativity. As we heard at this conference from
Claes Uggla, the conjecture implies that near space-like
singularities (such as the big bang) the homogeneity approximation
becomes successively better (because `space derivative terms' can
be neglected compared to `time derivative terms'). Therefore,
lessons from quantum theory in the homogeneous context may be more
potent than one might a priori imagine. The second point is
illustrated by the hydrogen atom. Suppose, hypothetically, we had
no experimental data on atomic spectra to assist us, nor a theory
of hydrogen atom but knew that the charged-particle-photon systems
should be treated using quantum electrodynamics. Suppose Dirac
then came up with his solution of the bound state problem for an
electron in an external Coulomb field. Since all but a finite
number of the field degrees of freedom are frozen in this model, a
priori one might have thought that the solution would be a poor
representation of the physical hydrogen atom in which true quantum
fluctuations can excite any and all of the frozen degrees of
freedom. A priori this is a perfectly reasonable concern but we
know that it is misplaced in the real world. Neither of these two
points is compelling but together they suggest that the
qualitative features of the symmetry reduced analysis could well
turn out to be robust.

Finally, the symmetry reduced models are generally useful in
providing intuition and guidance. For example, detailed
analytical/numerical studies in loop quantum cosmology \cite{aps}
have led to new insights for constructing the physical sector of
the theory, forced us to abandon naive dynamics and revealed
interesting physics in the Planck domain. In this respect these
calculations are somewhat analogous to checking AdS/CFT conjecture
in symmetric situations, e.g. the Penrose limit. Furthermore, FRW
cosmologies are not tailored just to make the problem tractable,
but are of direct physical interest.

\section{Assessment and comparison}
\label{s2.4}

\b \emph{11. If a definite Hamiltonian constraint has not yet
emerged in LQG, could one not say that there has been little
progress since the Wheeler-DeWitt geometrodynamics?}\\

It is perhaps clarifying to compare the situation with that of
string theory. There, the perturbation series is known to diverge
---furthermore it does so uncontrollably because it is not even
Borel summable. So, to any physical question ---such as the value
graviton-graviton scattering amplitude---  the full answer in
perturbative string theory is \emph{infinite}. Now, one may say
that this also happens in QED. But there we know the theory is
incomplete and we should not trust its predictions beyond a
certain number of terms in perturbation theory. But a theory which
claims to be complete can not take such a refuge. Therefore, the
standard belief in string theory is that the infinite answer of
the perturbation theory is incorrect because non perturbative
effects are crucial. This was realized almost two decades ago but
we still only have a skeleton of the candidate, non-perturbative
$M$ theory. So, the analogous question in string theory would be:
could one not say that there has been little progress since
supergravity?

I think that in both cases there is considerable incompleteness and
diversity of ideas on how to address it. \emph{But there has also
been considerable progress: In both cases, internally consistent
scenarios have emerged and special cases have been well-understood.}

Here are a few examples from LQG:\\
i) There is a  strong uniqueness theorem for the kinematical framework;\\
ii) We have a fairly  good understanding of the geometry of quantum
horizons in equilibrium and a statistical mechanical derivation of
entropy of astro-physically realistic black holes;\\
iii) Spin-foam models have led to a fertile approach to obtaining
the graviton propagator, effective low energy theories \emph{and}
probing non-perturbative aspects of Yang-Mills theories \cite{ym}; and,\\
iv) Dynamical ideas have been implemented \emph{in detail} in
mini-superspaces and have led to physically desirable results,
including singularity resolutions.\\ \medskip

\b {\emph{12. Are LQG and string theory two versions of the same
theory \emph{or} are they mutually incompatible?}\\

My personal answer is: Neither! Certainly, the two are not
equivalent in the sense that the Pauli's algebraic treatment of
the hydrogen atom is equivalent to Schr\"odinger's treatment based
on wave functions. Both LQG and string theory are very incomplete
and they start out with very different basic assumptions. These
differences are important but, in my view, they make the two
attempts complementary rather than incompatible. In particular,
LQG has taught lessons on implementation of background
independence and the use of quantum geometry in the resolution of
realistic singularities. String theory has provided a
qualitatively new strategy for unification, well developed
perturbative treatments and valuable experience with effective low
energy actions. All these are likely to be useful in our search
for a viable quantum gravity theory.

Interchange of ideas has been minimal so far because not only do
the two approaches have different starting points but they use
very different mathematical frameworks, making translations
non-trivial. For example, Hilbert spaces with `polymer-like' spin
network states and operators with discrete eigenvalues feature
prominently in LQG. String theory, on the other hand, tends to use
path integrals and continuum conformal field theories. Such
differences in mathematical frameworks naturally lead each
community to address a set of questions not easily accessible to
the other. For example, because of background independence, there
are no classical fields at all in the fundamental description of
LQG. Therefore, effective low energy actions which play an
important role in string theory are generally difficult to
construct. Reciprocally, the resolution of the FRW big-bang
singularity of loop quantum cosmology is difficult to recover in
string theory. Unfortunately, because of these deep differences of
emphasis and language, misconceptions can arise rather easily. But
this variety is also very good. For, as Feynman emphasized during
a lecture at CERN on his way back from Stockholm:
\begin{quote}

{\it ``It is very important that we do not all follow the same
fashion... It's necessary to increase the amount of variety .... the
only way to do it is to implore you few guys to take a risk ...'' }

\end{quote}

\section*{Acknowledgements}

I am grateful to John Barrett, Martin Bojowald, Alex Corichi,
Laurent Friedel, Jerzy Lewandowski, Kirill Krasnov, Jorge Pullin,
Carlo Rovelli, Lee Smolin and Thomas Thiemann for illuminating
discussions and correspondence. I would also like to thank Gary
Horowitz, Don Marolf, Hermann Nicolai, Joe Polchinski, and Stephen
Shenker for their questions and comments.  This work was supported
in part by the NSF grant PHY04-56913, the Alexander von Humboldt
Foundation, the Kramers Chair program of the University of Utrecht,
and the Eberly research funds of Penn State.


\begin{thebibliography}{99}

\vfill\break

\bibitem{lost} J.~Lewandowski, A.~Okolow, H.~Sahlmann and
    T.~Thiemann, {Uniqueness of diffeomorphism invariant states
    onholonomy flux algebras},  Comm. Math. Phys. \textbf{267},
    703-733 (2006) \texttt{arXiv:gr-qc/0504147};\\
C.~Fleishchack, {Representations of the Weyl algebra in quantum
geometry}, \texttt{arXiv:math-ph/0407006}.

\bibitem{aps} A.~Ashtekar, T.~Pawlowski and P.~Singh, {Quantum
    nature of the big bang}, Phys. Rev. Lett. \textbf{96}, 141301
    (2006), \texttt{arXiv:gr-qc/0602086};\\
{Quantum nature of the big bang: An analytical and numerical
investigation I}, Phys. Rev. {\bf D73}, 124038, \texttt{arXiv:gr-qc/0604013};\\
{Quantum nature of the big bang: Improved dynamics}, Phys. Rev.
{\bf D74}, 084003, \texttt{arXiv:gr-qc/0604013}; \\
A.~Ashtekar, T.~Pawlowski, P.~Singh and K.~Vandersloot, {Loop
quantum cosmology of $k$=1 FRW models}, Phys. Rev. \textbf{D75},
0240035 (2007), \texttt{arXiv:gr-qc/0612104};\\
L.~Szulc, W.~Kaminski, J.~Lewandowski, Closed FRW model in loop
quantum cosmology, \texttt{arXiv:gr-qc/0612101};\\
J.~Brunnemann and T.~Thiemann, On (cosmological) singularity
avoidance in loop quantum gravity, Class. Quant. Grav. \textbf{23}
1395-1428 (2006).

\bibitem{fl} L.~Friedel and E.R.~Livine, Effective 3-d quantum
    gravity and non-commutative field theory, Phys. Rev. Lett.
    \textbf{96}, 221301 (2006), \texttt{rXiv:hep-th/0512113};\\
Ponzano-Regge model revisited III: Feynman diagrams and effective
field theory, Class. Quant. Grav. \textbf{23}, 2021--2062 (2006),
\texttt{rXiv:hep-th/0502106}.

\bibitem{propagator} C.~Rovelli, Graviton propagator from
    background-independent quantum gravity, Phys. Rev. Lett.
    \textbf{97}, 151301 (2006), \texttt{arXiv:gr-qc/0508124};\\
E.~Bianchi, L.~Modesto, C.~Rovelli, S.~Speziale, Graviton propagator
in loop quantum gravity, Class. Quantum Grav.
\textbf{23}, 6989-7028 (2006), \texttt{arXiv:gr-qc/0604044};\\
E.~R.~Livine and S.~Speziale, Group Integral Techniques for the
Spinfoam Graviton Propagator, \texttt{arXiv:gr-qc/0608131}.

\bibitem{ao} A.~Okolow, Representations of Quantum Geometry, PhD thesis,
University of Warsaw (2005).

\bibitem{master} T.~Thiemann, The Phoenix Project: Master Constraint
    Programme for Loop Quantum Gravity, Class. Quant. Grav. \textbf{23},
    2211-2248 (2006), \texttt{gr-qc/0305080};\\
Quantum Spin Dynamics VIII. The Master Constraint,
Class. Quant. Grav. \textbf{23}, 2249-2266 (2006)\\
B.~Dittrich and T.~Thiemann, Testing the Master Constraint
Programme for Loop Quantum Gravity I. General Framework, Class.
Quant. Grav. \textbf{23}, 1025-1066 (2006), \texttt{gr-qc/0411138};\\
Testing the Master Constraint Programme for Loop Quantum Gravity
II. Finite Dimensional Systems, Class. Quant. Grav. \textbf{23},
1067-1088 (2006), \texttt{gr-qc/0411139};\\
Testing the Master Constraint Programme for Loop Quantum Gravity
III. SL(2,R) Models, Class. Quant. Grav. \textbf{23},
1089-1120 (2006), \texttt{gr-qc/0411140};\\
Testing the Master Constraint Programme for Loop Quantum Gravity
IV. Free Field Theories, Class. Quant. Grav. \textbf{23},
1121-1142 (2006), \texttt{gr-qc/0411141}.

\bibitem{aqg} K.~Giesel and T.~Thiemann, Algebraic Quantum Gravity
    (AQG)I. Conceptual Setup, \texttt{gr-qc/0607099};\\ Algebraic Quantum
    Gravity (AQG) II. Semiclassical Analysis, \texttt{gr-qc/0607101};\\
Algebraic Quantum Gravity (AQG) III. Semiclassical Perturbation
Theory, \texttt{gr-qc/0607101}.

\bibitem{madrid} J.F.~Barbero G., I.~Garay, E.J.S.~Villaseñor,
Exact quantization of Einstein-Rosen waves coupled to massless
scalar matter, Phys. Rev. Lett. \textbf{95}, 050501 (2005),
\texttt{arXiv:gr-qc/0506093};\\
Probing quantized Einstein-Rosen waves with massless scalar
matter, Phys. Rev. \textbf{D74} 044004 (2006) \texttt{gr-qc/0607053};\\
J.F.~Barbero G., G.A.~Mena Marugán, E.J.S.~Villaseñor, Asymptotics
of Regulated Field Commutators for Einstein-Rosen Waves, J. Math.
Phys.\textbf{46} 062306 (2005), \texttt{gr-qc/0412028};\\
Particles and vacuum for perturbative and non-perturbative
Einstein-Rosen gravity, Phys. Rev. \textbf{D70}, 044028 (2004),
\texttt{gr-qc/0406087};\\
Quantum cylindrical waves and sigma models, Int.J.Mod.Phys.
\textbf{D13}, 1119-1128 (2004); \texttt{gr-qc/0402096}.

\bibitem{matter} T.~Thiemann, QSD V : Quantum gravity as the natural
    regulator of matter quantum field theories, Class. Quant. Grav.
    \textbf{15}, 1281-1314 (1998), \texttt{arXiv:gr-qc/9705019}.

\bibitem{ab} A.~Ashtekar and M.~Bojowald, Black hole evaporation:
A paradigm, Class. Quant. Grav. \textbf{22}, 3349-3362 (2005);
{Quantum geometry and the Schwarzschild singularity}, Class.
Quant. Grav. {\bf 23} (2006) 391-411, \texttt{arXiv:gr-qc/0509075}\\
L.~Modesto, Loop quantum black hole, Class. Quant. Grav.
\textbf{23}, 5587-5602 (2006), \texttt{arXiv:gr-qc/0509078v2}.

\bibitem{mb} M.~Bojowald, {Absence of singularity in loop quantum
    cosmology}, Phys. Rev. Lett. \textbf{86}, 5227-5230 (2001),
    \texttt{arXiv:gr-qc/0102069};\\
{Isotropic loop quantum cosmology}, Class. Quantum. Grav.
\textbf{19}, 2717-2741 (2002), \texttt{arXiv:gr-qc/0202077};\\
A.~Ashtekar, M.~Bojowald, J.~Lewandowski, {Mathematical structure of
loop quantum cosmology}, Adv. Theo. Math. Phys. \textbf{7}, 233-268
(2003), \texttt{gr-qc/0304074}.
\bibitem{mbrev} M.~Bojowald, {Loop quantum cosmology}, Liv. Rev.
    Rel. \textbf{8}, 11 (2005), \texttt{arXiv:gr-qc/0601085}.

\bibitem{alrev} A.~Ashtekar and J.~Lewandowski, {Background
    independent quantum gravity: A status report}, Class. Quant.
    Grav. {\bf 21}, R53-R152 (2004), \texttt{arXiv:gr-qc/0404018}.

\bibitem{crbook} C.~Rovelli {\em Quantum Gravity}, (CUP, Cambridge,
    2004).

\bibitem{ttbook} T.~Thiemann, {\em Introduction to Modern Canonical
    Quantum General Relativity} (CUP, Cambridge, at press)
\\
Loop Quantum Gravity: An Inside View, \texttt{hep-th/0608210}.

\bibitem{aljb} A.~Ashtekar and C.~J.~Isham, Representation of the holonomy
algebras of gravity and non-Abelian gauge theories, Class. Quant.
Grav. {\bf 9}, 1433-1467 (1992);\\
A.~Ashtekar and J.~Lewandowski, Representation theory of analytic
holonomy algebras, in {\sl Knots and Quantum Gravity}, edited by
J.\ C.\ Baez, Oxford U.\ Press, Oxford, (1994),
\texttt{arXiv:gr-qc/9311010};\\
Projective techniques and functional integration, Jour. Math.
Phys. {\bf 36}, 2170-2191 (1995), \texttt{arXiv:gr-qc/9411046}\\
Differential geometry on the space of connections using projective
techniques, Jour. Geo. \& Phys. {\bf 17}, 191-230 (1995);
\texttt{arXiv:gr-qc/9412073};\\
J.~C.~Baez, Generalized measures in gauge theory, Lett. Math.
Phys. {\bf 31}, 213-223 (1994),  \texttt{arXiv:hep-th/9310201}; \\
Diffeomorphism-invariant generalized measures on the space of
connections modulo gauge transformations, In Proceedings of the
Conference on Quantum Topology, ed. David N. Yetter, World
Scientific Press, Singapore (1994) pp. 21-43;\\
D.~Marolf and J.~Mour\~ao, On the support of the
Ashtekar-Lewandowski measure, Commun. Math. Phys.{\bf 170} ,
583-606 (1995), \texttt{arXiv:hep-th/9403112}.

\bibitem{spinnet} C.~Rovelli and L.~Smolin, Spin networks and
quantum gravity, Phys. Rev. {\bf D52}, 5743-5759 (1995),
\texttt{arXiv:gr-qc/9505006};\\
J.~Baez, Spin network states in gauge theory, Adv. Math.
\textbf{117}, 253-272 (1996), \texttt{gr-qc/9411007}.

\bibitem{bd} B.~S.~DeWitt, Quantum theory of gravity. I. The
    canonical theory, Phys. Rev. \textbf{160} 1113-1148 (1967).

\bibitem{cm} C.~W.~Misner, Mixmaster universe, Phys. Rev. Lett.
\textbf{22}, 1071-1074 (1969);\\
Minisuperspace, in: \textit{Magic without Magic: John Archibald
Wheeler; a collection of essays in honor of his sixtieth birthday},
edited by J.~R.~Klauder (W. H. Freeman, San Francisco, 1972).

\bibitem{gu} D.~Green and W.~Unruh, Difficulties with recollapsing
    models in closed isotropic loop quantum cosmology, Phys. Rev.
    \textbf{ D70}, 103502 (2004), \texttt{arXiv:gr-qc/04-0074}.

\bibitem{st} H.~Sahlmann and T.~Thiemann, Towards the QFT on curved
spacetime limit of QGR. I: A general scheme, Class. Quant. Grav.
\textbf{23}, 867-908 (2006), \texttt{arXiv:gr-qc/0207030} ;\\
Towards the QFT on curved spacetime limit of QGR. II: A concrete
implementation, Quant. Grav. \textbf{23}, 909-954 (2006),
\texttt{arXiv:gr-qc/0207031}

\bibitem{spinfoams}M.~P.~Reisenberger and C.~Rovelli,
'Sum over surfaces' form of loop quantum gravity, Phys. Rev.
\textbf{D56}, 3490-3508 (1997), \texttt{arXiv:gr-qc/9612035};\\
J.~C.~Baez, Spin foam models, Class. Quant. Grav.
\textbf{15} 1827-1858 (1998), \texttt{arXiv:gr-qc/9709052};\\
J.~W.~Barrett and L.~Crane, A Lorentzian signature model for
quantum general relativity, Class. Quant. Grav.
\textbf{17},3101-3118 (2000), \texttt{arXiv:gr-qc/9904025}.

\bibitem{snyder} H.~S.~Snyder, Quantized space-time, Phys. Rev.
    \textbf{71}, 38-41 (1947).

\bibitem{jk}J.~Klauder, Fundamentals of quantum gravity, \texttt{
arXiv:gr-qc/0612168};\\
Overview of affine quantum gravity, Int. J. Geom. Meth. Mod. Phys.
\textbf{3}, 81-94(2006) \texttt{arXiv:gr-qc/0507113}.

\bibitem{higherj} K.~Vandersloot, On the Hamiltonian constraint of
loop quantum cosmology, Phys. Rev. D\textbf{71} 103506 (2005),
\texttt{gr-qc/0502082};\\
Ph.D. Dissertation, submitted to The Pennsylvania State University
(2006);\\
A.~Perez, {On the regularization ambiguities in loop quantum
gravity}, Phys.Rev. D {\bf 73} 044007 (2006),
\texttt{arXiv:gr-qc/0509118}.

\bibitem{2+1} A.~Perez, Introduction to loop quantum gravity and
spin foams, Lectures presented at the II International Conference
of Fundamental Interactions, Pedra Azul, Brazil, June 2004,
\texttt{arXiv:gr-qc/0409061};\\
A.~Ashtekar, V.~Husain, C.~Rovelli, J.~Samuel and L.~Smolin, 2+1
quantum gravity as a toy model for the 3+1 theory, Class. Quant.
Grav. \textbf{6}, L185-L193 (1989);\\
L.~Freidel and D.~Louapre, Ponzano-Regge model revisited I: Gauge
fixing, observables and interacting spinning particles, Class.
Quant. Grav. \textbf{21}, 5685-5726 (2004),
\texttt{arXiv:hep-th/0401076}.

\bibitem{cdt}J.~Ambjorn, J.~Jurkiewicz, and R.~Loll, A non-perturbative
Lorentzian path integral for gravity, Phys. Rev. Lett.
\textbf{85}, 924-927 (2000), \texttt{hep-th/0002050};\\
Quantum gravity, or The art of building spacetime,
\texttt{hep-th/0604212}

\bibitem{reuter}O~Luscher and M.~Reuter, Asymptotic safety in
quantum Einstein gravity: nonperturbative renormalizability and
fractal spacetime structure, \texttt{arXiv:hep-th/0511260};\\
 Flow Equation of Quantum Einstein Gravity in a Higher-Derivative
 Truncation, \texttt{arXiv:hep-th/0205062}.

\bibitem{crss}C.~Rovelli and S.~Speziale, Reconcile Planck scale
    diescreteness and the Lorentz-Fitzgerald contraction, Phys. Rev.
    \textbf{D67}, 064019 (2003), \texttt{arXiv:gr-qc/0205108}.

\bibitem{lv}D.~Mattingly,  Modern tests of Lorentz invariance, Living
Rev. Rel. \textbf{8} 5, (2005) 5, \texttt{ arXiv:gr-qc/0502097};\\
J.~Collins, A.~Perez and D.~Sundarsky, L.~Urritia and H.~Vucetich,
Lorentz invariance and quantum gravity: an additional fine-tuning
problem?, Phys. Rev. Lett. \textbf{93}, 191301 (2004).

\bibitem{abk} A.~Ashtekar, J.~Baez and K.~Krasnov, Quantum geometry
of isolated horizons and black hole entropy,  Adv. Theor. Math.
Phys. \textbf{4}, 1-94 (2000), \texttt{arXiv:gr-qc/0005126}.

\bibitem{ym} L.~Freidel, R.~G.~Leigh and D.~Minic, Towards a solution of
pure Yang-Mills theory in 3+1 dimensions, Phys. Lett.
\textbf{B641}, 105-111 (2006), \texttt{arXiv:hep-th/0604184};\\
L.~Friedel, On pure Yang-Mills theory in 3+1 dimensions:
Hamiltonian, vacuum and gauge invariant variables,
\texttt{arXiv:hep-th/0604185v2};\\
F.~Conrady, Analytic derivation of dual gluons and monopoles from
SU(2) lattice Yang-Mills theory. II. Spin foam representation,
\texttt{arXiv:hep-th/0610237}.

\bibitem{rf} J.~Gleick, \textit{Genius} (Patheon, 1992), page 382.

\end{thebibliography}
\end{document}